\documentclass[10pt,prc,aps,psfig,twocolumn]{revtex4}
\usepackage{graphicx}
\tighten         
\begin{document}
\draft
\title{Comparing neutron star predictions by various microscopic models                              
}
\author{F. Sammarruca and P. Liu}
\affiliation{Physics Department, University of Idaho, Moscow, ID 83844, U.S.A}
\date{\today}

\email[F. Sammarruca: ]{fsammarr@uidaho.edu}

\begin{abstract}
We calculate several neutron star properties, for static and/or rotating stars, using 
equations of state based on different microscopic models. These include our 
Dirac-Brueckner-Hartree-Fock model and others derived from the non-relativistic
Brueckner-Hartree-Fock approach implemented with microscopic three-body forces.
The model dependence is discussed. 
\end{abstract}

\maketitle

\section{Introduction}                                                  

The determination of the equation 
of state (EoS) 
of highly compressed and/or hot matter is one of the most complex problems in    
nuclear physics. 
In terrestrial laboratories, relativistic
heavy-ion collisions are the best tool to produce hot and dense matter. Concerning astrophysical systems,
neutron stars are known to be stable configurations containing the most dense form of matter found in the universe.
They are therefore unique laboratories to study the properties of highly compressed (cold) matter. 
Furthermore, the possibility of studying the structure of neutron stars via gravitational waves \cite{GW} 
makes these exotic objects even more exciting. 

More generally, 
the study of neutron-rich systems (from the lowest to the highest densities) has widespread
impact, reaching from the physics of exotic nuclei to nuclear astrophysics. 
With the Facility for Rare Isotope Beams (FRIB) recently approved for design and construction at 
Michigan State University, these studies 
 become particularly important and timely.                                     
Partnership between nuclear physics and astrophysics will play a crucial role in 
advancing our knowledge of neutron-rich matter and its equation of state.

Clearly, a rich (on-going or future) experimental program must be accompanied by parallel theoretical
effort. Present nuclear matter calculations are typically performed either within the mean field approach or the
 microscopic one. 
Mean field models, relativistic and non-relativistic, are popular \cite{Sk,RMF}, but   
in our opinion {\it ab initio} approaches are best in order to get deeper      
physical insight. By {\it ab initio}, we mean that the starting point is a realistic
two-body potential, possibly complemented by three-body forces. 

The purpose of this paper is to compare several neutron star properties  predicted by different microscopic models.
As constraints promise to become more stringent, it is important to understand and compare how the nature of the various predictions
is related to the features of each model. In microscopic approaches, the 
tight connection with the underlying two-body potential will then facilitate the physical understanding 
in terms of the characteristics of the nuclear force and its behavior in the medium. 

In the next section, we review the main aspects of the {\it ab initio} approach. 
The main differences between our approach (the Dirac-Brueckner-Hartree-Fock method) and the 
other models used in our comparison (the conventional 
Brueckner-Hartree-Fock method together with microscopic three-body forces) 
are revisited and discussed. 

\section{The ``ab initio" approach}                                                

\subsection{The two-body framework} 
Our present knowledge of the nuclear force is the result of decades of
struggle \cite{Mac89}. After the development of QCD and the understanding of its symmetries,  
chiral effective theories \cite{chi} became popular as a way to respect the       
symmetries of QCD while keeping the degrees of freedom (nucleons and pions) typical of low-energy nuclear physics. However, 
chiral perturbation theory (ChPT)
has definite limitations as far as the range of allowed momenta is concerned. 
For the purpose of applications in dense matter, where higher and higher momenta become involved     
with increasing Fermi momentum, ChPT is inappropriate. 
A relativistic, meson-theoretic model is the better choice.                    

The one-boson-exchange (OBE) model has proven very successful in describing nucleon-nucleon (NN) data in free space 
and has a good theoretical foundation. 
Among the many available OBE potentials (some being part of the ``high-precision generation"              
\cite{pot1,pot2,pot3}), 
we seek a momentum-space potential developed within a relativistic scattering equation, such as the 
one obtained through the Thompson three-dimensional reduction of the Bethe-Salpeter equation.
Furthermore, we require a potential that uses 
the pseudovector coupling for the interaction of nucleons with pseudoscalar mesons. 
With this in mind, 
as well as the requirement of a good description of NN data, 
Bonn B \cite{Mac89} has been our standard choice. As is well known, the NN potential model dependence
of nuclear matter predictions is not negligible. The saturation points obtained with different NN potentials
move along the famous ``Coester band" depending on the strength of the tensor force, with the weakest tensor
force corresponding to the largest attraction. For the same reason (that is, the role of the tensor force in  
nuclear matter), 
the potential model dependence is strongly reduced in pure (or nearly pure) neutron matter, due to the  
absence of isospin-zero partial waves. 

Already when QCD (and its symmetries) were unknown, it was observed that the contribution from the
nucleon-antinucleon pair diagram, Fig.~1, is unreasonably large when the pseudoscalar (ps) coupling is used, 
leading to very large pion-nucleon scattering lengths \cite{GB79}.                                            
We recall that the Lagrangian density for pseudoscalar coupling of the nucleon field ($\psi$) with the  pseudoscalar meson
field ($\phi$) is 
\begin{equation}
{\cal L}_{ps} = -ig_{ps}\bar {\psi} \gamma _5 \psi \phi. 
\end{equation} 
On the other hand, the same contribution (Fig.~1) 
is heavily suppressed by the pseudovector (pv) coupling (a mechanism which
became known as ``pair suppression"). The reason for the suppression is the presence of the 
covariant derivative                                                                                     
at the pseudovector vertex,                                                  
\begin{equation}
{\cal L}_{pv} = \frac{f_{ps}}{m_{ps}}{\bar \psi}  \gamma _5 \gamma^{\mu}\psi \partial_{\mu} \phi, 
\end{equation} 
which suppresses the vertex for low momenta and, thus, 
 explains the small value of the pion-nucleon
scattering length at threshold \cite{GB79}. 
Considerations based on chiral symmetry \cite{GB79} can further motivate 
the choice of the pseudovector coupling.                          
We will come back to this point in the next subsection. 

The most important aspect of the ``{\it ab initio}" approach is that the only free parameters of the
model (namely, the parameters of the NN potential)                                               
are determined by the fit to the free-space data and never readjusted in the medium. In other
words, the model parameters are tightly constrained and the calculation in the medium is 
parameter free. 
The presence of free parameters in the medium would generate effects and sensitivities which can be very large and hard to
control. 

\begin{figure}
\centering            
\vspace*{-4.0cm}
\hspace*{-2.0cm}
\scalebox{0.8}{\includegraphics{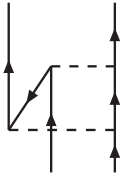}}
\vspace*{-17.0cm}
\caption{Contribution to the NN interaction from virtual pair excitation.                   
Upward- and downward-pointing arrows represent nucleons and antinucleons, respectively.
Dashed lines denote mesons.                            
} 
\label{one}
\end{figure}
\subsection{The many-body framework: Brueckner theory, three-body forces, and relativity} 

Excellent reviews of Brueckner theory have been written which we can refer the reader to 
(see \cite{Mac89} and references therein). 
Here, we begin by defining the contributions that are retained in our calculation. Those are 
the lowest order contribution to the Brueckner series (two-hole lines) and the corresponding
exchange diagram. 
With the G-matrix as the effective interaction, this amounts to including particle-particle
(that is, short-range) correlations, which are absolutely essential to even approach a realistic
description of nuclear matter properties. 
Three-nucleon correlations have been shown to be small if the continuous choice is adopted 
for the single-particle potential \cite{Baldo98}. 

The issue of three-body forces (TBF), of course, remains to be discussed. 
In Fig.~2 we show a TBF originating from virtual excitation of a nucleon-antinucleon pair, 
known as ``Z-diagram". Notice that the observations from the previous section ensure that the corresponding diagram
at the two-body level, Fig.~1, is small with pv coupling. 
At this point, it is useful 
to recall the main feature of the Dirac-Brueckner-Hartree-Fock 
(DBHF) method, as that turns out to be closely related to 
the TBF depicted in Fig.~2. In the DBHF approach, one describes the positive energy solutions
of the Dirac equation in the medium as 
\begin{equation}
u^*(p,\lambda) = \left (\frac{E^*_p+m^*}{2m^*}\right )^{1/2}
\left( \begin{array}{c}
 {\bf 1} \\
\frac{\sigma \cdot \vec {p}}{E^*_p+m^*} 
\end{array} 
\right) \;
\chi_{\lambda},
\end{equation}
where the effective mass is given by $m^* = m+U_S$, with $U_S$ an attractive scalar potential.
It turns out that both the description of a single-nucleon via Eq.~(3) and the evaluation of the 
Z-diagram, Fig.~2, generate a repulsive effect on the energy/particle in symmetric nuclear matter which depends on the density approximately
as 
\begin{equation}
\Delta E \propto  \left (\frac{\rho}{\rho_0}\right )^{8/3}, 
\end {equation}
and provides the saturating mechanism missing from conventional Brueckner calculations. 
Brown showed that the bulk of this effect can be obtained as a lowest order (in $p^2/m$) relativistic correction
to the single-particle propagation \cite{GB87}. 

\begin{figure}[!t] 
\centering         
\vspace*{-4.0cm}
\hspace*{-2.0cm}
\scalebox{0.8}{\includegraphics{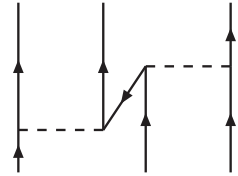}}
\vspace*{-17.0cm}
\caption{Three-body force due to virtual pair excitation. Conventions as in the previous figure.
} 
\label{two}
\end{figure}

The approximate equivalence of the effective-mass spinor description and the contribution from the Z-diagram 
has a simple intuitive explanation in the observation 
that Eq.~(3), like any other solution of the Dirac equation,
can be written as a combination of positive and negative energy solutions. On the other hand, the ``nucleon" in the 
middle of the Z-diagram, Fig.~2, is precisely a superposition of positive and negative energy states. 
In summary, the DBHF method effectively takes into account a particular class of 
TBF, which are crucial for nuclear matter saturation, 
but does not include other TBF.                                                                 

Concerning other, more popular, three-body forces,                                
Fig.~3 shows the TBF that is included in essentially all TBF models, regardless
other components; it is the Fujita-Miyazawa TBF \cite{FM}.                                          
With the addition of contributions from $\pi N$ S-waves, one ends up with the 
well-known Tucson-Melbourne TBF \cite{TM}. The microscopic TBF of Ref.~\cite{Catania} include 
contributions from excitations of the Roper resonance (P$_{11}$ isobar) as well. 

Now, if diagrams such as the one shown on the left-hand side of  Fig.~3 are included, 
consistency requires that medium modifications at the corresponding two-body level are also included, 
that is, the diagram on the right-hand side of Fig.~3 should be present and properly medium modified. 
Large cancellations then take place, a fact that was brought up a long time ago \cite{DMF}.                       
When the two-body sector is handled via OBE diagrams, the two-pion exchange is 
effectively incorporated through the $\sigma$ ``meson", which 
cannot generate the (large) medium effects (dispersion and Pauli blocking on $\Delta$
intermediate states) required by the consistency arguments above.

\begin{figure}
\begin{center}
\vspace*{-4.0cm}
\hspace*{-2.0cm}
\scalebox{0.8}{\includegraphics{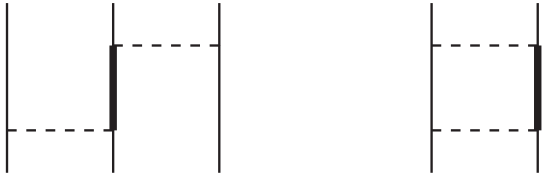}}
\vspace*{-17.0cm}
\caption{Left: three-body force arising from $\Delta$-isobar excitation (thick line). 
Right: two-meson exchange contribution to the NN interaction involving 
$\Delta$-isobar excitation.
} 
\label{three}
\end{center}
\end{figure}

To contrast this point of view, 
we will take as the other elements of our comparison the EoS's from the microscopic approach 
of Ref.~\cite{cat}. There (and in previous work by the authors \cite{Catania}), 
the Brueckner-Hartree-Fock (BHF) formalism is employed along with microscopic three-body forces.        
However, in Ref.~\cite{Catania}
the meson-exchange TBF are constructed applying the same parameters
as used in the corresponding nucleon-nucleon (NN) potentials, which are: Argonne V18 (V18, \cite{V18}), Bonn B (BOB, \cite{Mac89}), Nijmegen 93 (N93, \cite{pot2}). 
The popular (but phenomenological) Urbana TBF (UIX, \cite{UIX}) is also utilized in Ref.~\cite{cat} along with 
the V18 potential. The parameters which are not specified by the NN potential are chosen according to
independent investigations \cite{Catania}. 
Convenient parametrizations in terms of simple analytic functions are given in all cases        
for the resulting EoS's. 
We will refer to this approach, generally, as ``BHF + TBF". 

  In a previous work \cite{SL09} we compared the 
 predictions of the neutron skin in $^{208}$Pb by these models and correlated them  
with differences in the slope of the symmetry energy. 
Model differences become larger at high-density and will naturally impact neutron star predictions. 

\section{Neutron stars: Results and discussion}
Stellar matter contains neutrons in $\beta$
equilibrium with protons, electrons, and muons. 
Our DBHF EoS for $\beta$-equilibrated matter is given in Ref.~\cite{Sam0806}. We have applied $\beta$-stability in the same way to the 
various models of        
 Ref.~\cite{cat} starting from the given parametrized versions of the respective symmetric matter and neutron matter EoS's.
At subnuclear densities, all the EoS's considered here are joined with the crustal equations of state from  the work of Harrison and Wheeler \cite{HW} (for energy densities between 10 and 10$^{11}$ g~cm$^{-3}$) and the work of Negele and Vautherin
\cite{NV} (for energy densities less than 1.7$\times$10$^{13}$g~cm$^{-3}$). 
The composition of the crust is crystalline, with light \cite{HW} or heavy \cite{NV} metals and electron gas.
The DBHF equation of state as applied in this work, including the crust, is given in the tables.  
All neutron star properties are
calculated using public software downloaded from the website {\it http://www.gravity.phys.uwm.edu/rns}. 

In Fig.~4, we show the mass-radius relation for a sequence of static neutron stars as predicted
by the various models listed above. All models besides DBHF share the same many-body
approach but differ in the two-body potential and TBF employed. The resulting differences can be 
much larger than those originating from the use of different many-body approaches. This can be seen by comparing
the DBHF and BOB curves, both employing the Bonn B interaction. Overall, the maximum masses range from 1.8$M_{\odot}$
(UIX) to 
2.5$M_{\odot}$ (BOB). Radii are less sensitive to the EoS and range between 10 and 12 km for all models under
consideration, DBHF or BHF+TBF. 
Concerning available constraints, an initial observation of a neutron-star-white dwarf binary system
suggested a neutron star mass (PSR J0751+1807) of 2.1$\pm$0.2$M_{\odot}$ \cite{Nice}. Such observation 
would imply a considerable constraint on the high-density behavior of the EoS. On the other hand,    
a dramatically reduced value             
of 1.3$\pm$0.2$M_{\odot}$                                          
was recently reported \cite{Piek08}, which does not invalidate any of the theoretical models under
consideration. 

The model dependence is shown in Fig.~5  for the case of rapidly rotating stars. 
The 716 Hz frequency corresponds to the most rapidly rotating pulsar, PSR J1748-2446 \cite{Hessels}, 
although recently an X-ray burst oscillation at a frequency of 1122 Hz has been reported \cite{Kaaret}
which may be due to the spin rate of a neutron star.
As expected, the maximum mass and the            
(equatorial) radius become larger with increasing rotational frequency.

Another bulk property of neutron stars is the moment of inertia, $I$. 
In Fig.~6, we show the moment of inertia at different rotational speeds (again, for all models), whereas in 
Fig.~7 we display the moment of inertia corresponding to the maximum mass at different
rotational frequencies.
These values are not in contraddiction with observations of the 
Crab nebula luminosity. From that, a lower bound on the moment of inertia was inferred to be 
$I \geq  $4-8 $\times$ 10$^{44}$ g cm$^2$, see Ref.~\cite{Weber} and references therein. 
The size of $I_{M_{max}}$ changes from model to model in line with the size of the maximum mass,
see Fig.~7. 
(We recall that the variations among radii from different EoS's are relatively mild). 

\begin{figure}[!t] 
\centering          
\includegraphics[totalheight=2.5in]{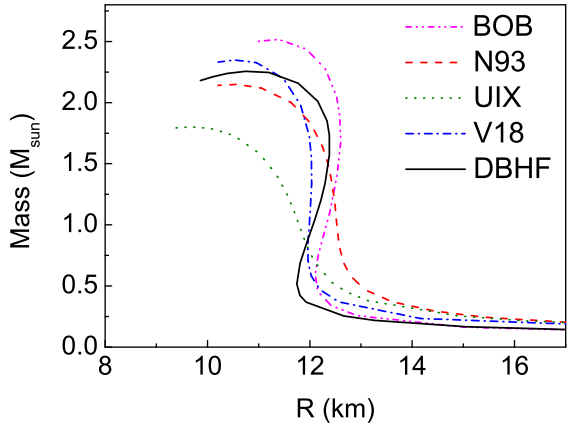} 
\vspace*{-0.5cm}
\caption{(color online) Static neutron star mass-radius relation for the models
considered in the text. 
} 
\label{four}
\end{figure}

\begin{figure}[!t] 
\centering          
\includegraphics[totalheight=1.8in]{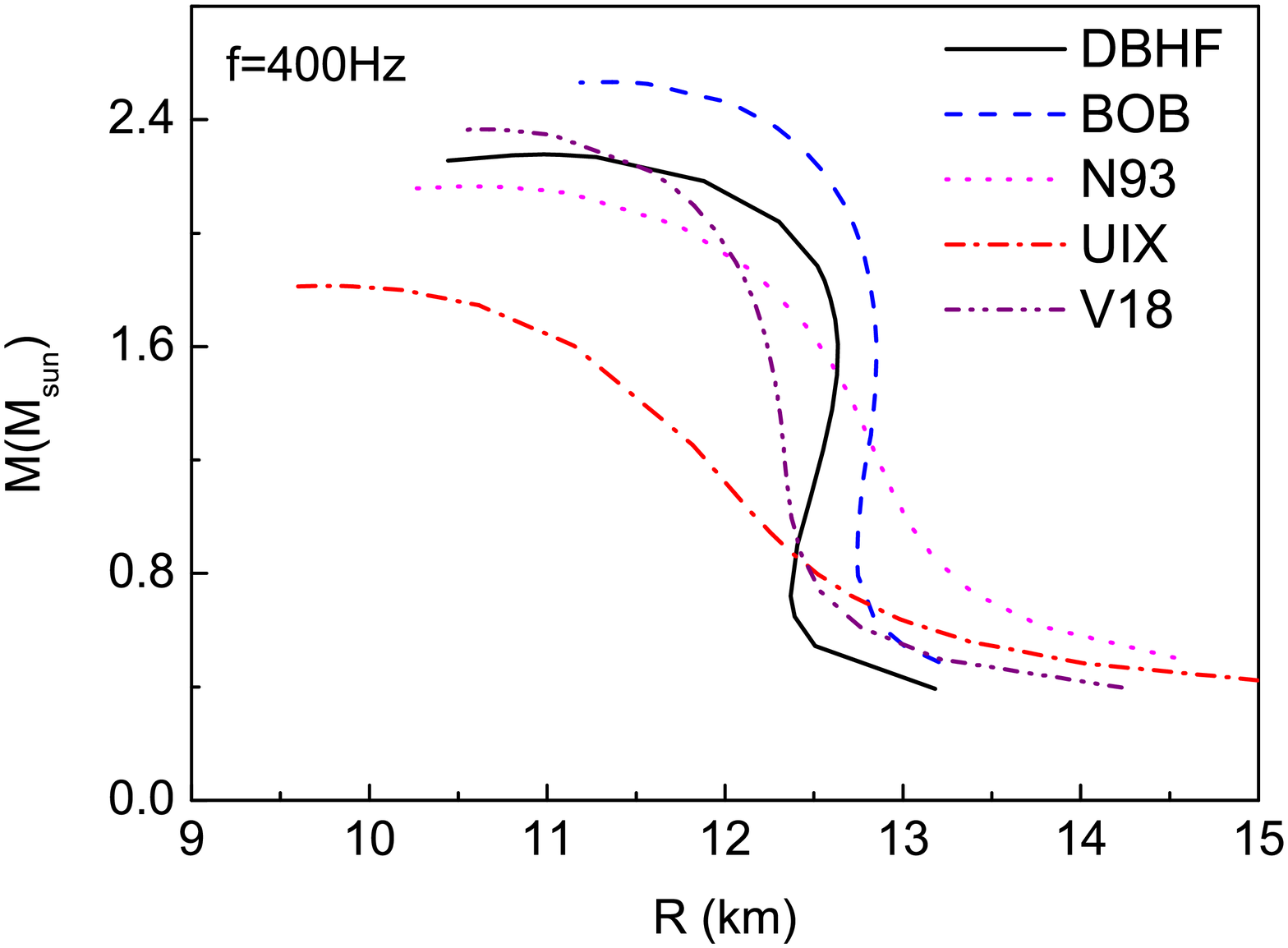}  
\includegraphics[totalheight=1.8in]{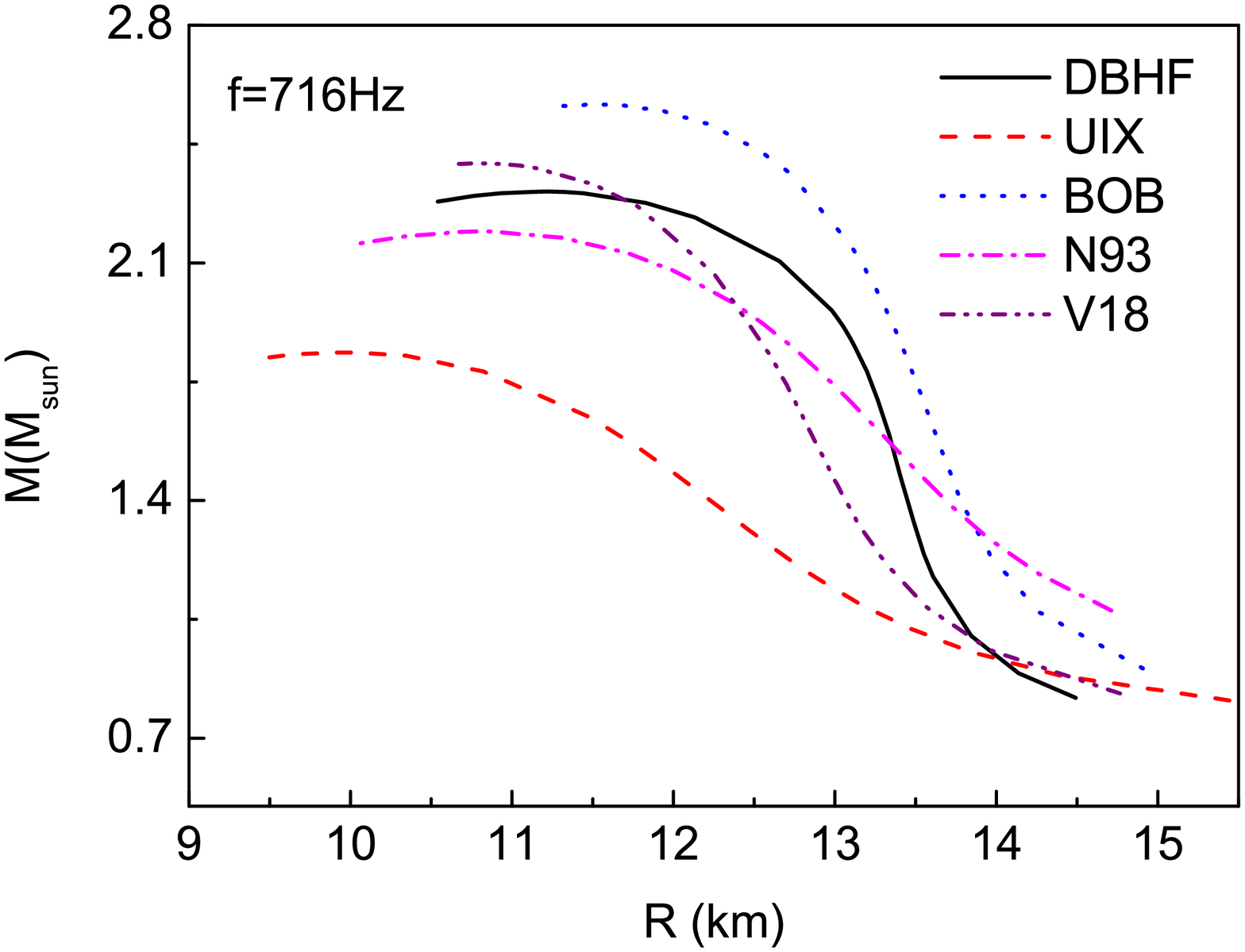}  
\includegraphics[totalheight=1.8in]{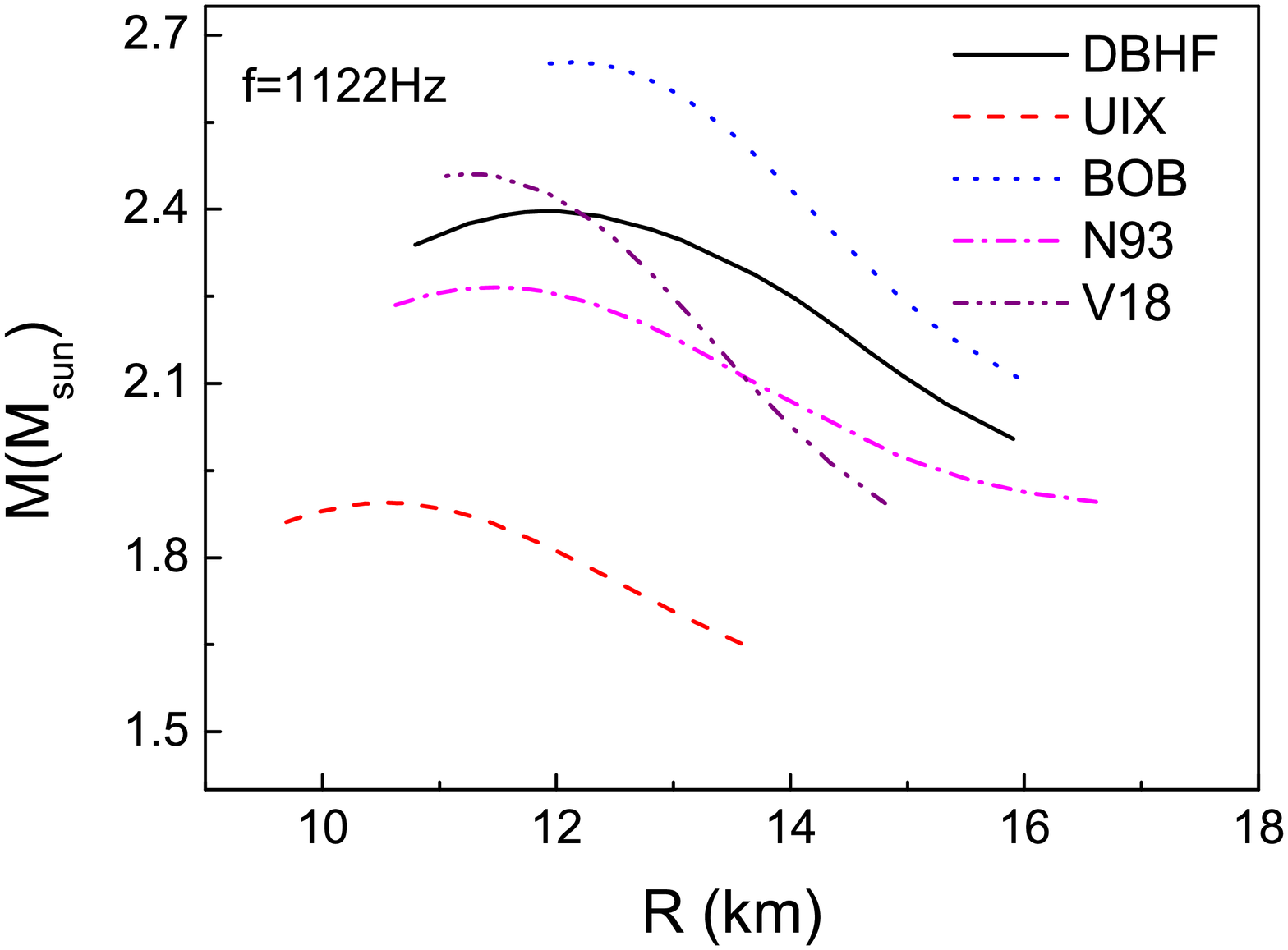}    
\vspace*{-0.5cm}
\caption{(color online)  Mass-radius relation for the models
considered in the text and for different rotational frequencies.                  
} 
\label{five}
\end{figure}

\begin{figure}[!t] 
\centering          
\includegraphics[totalheight=1.8in]{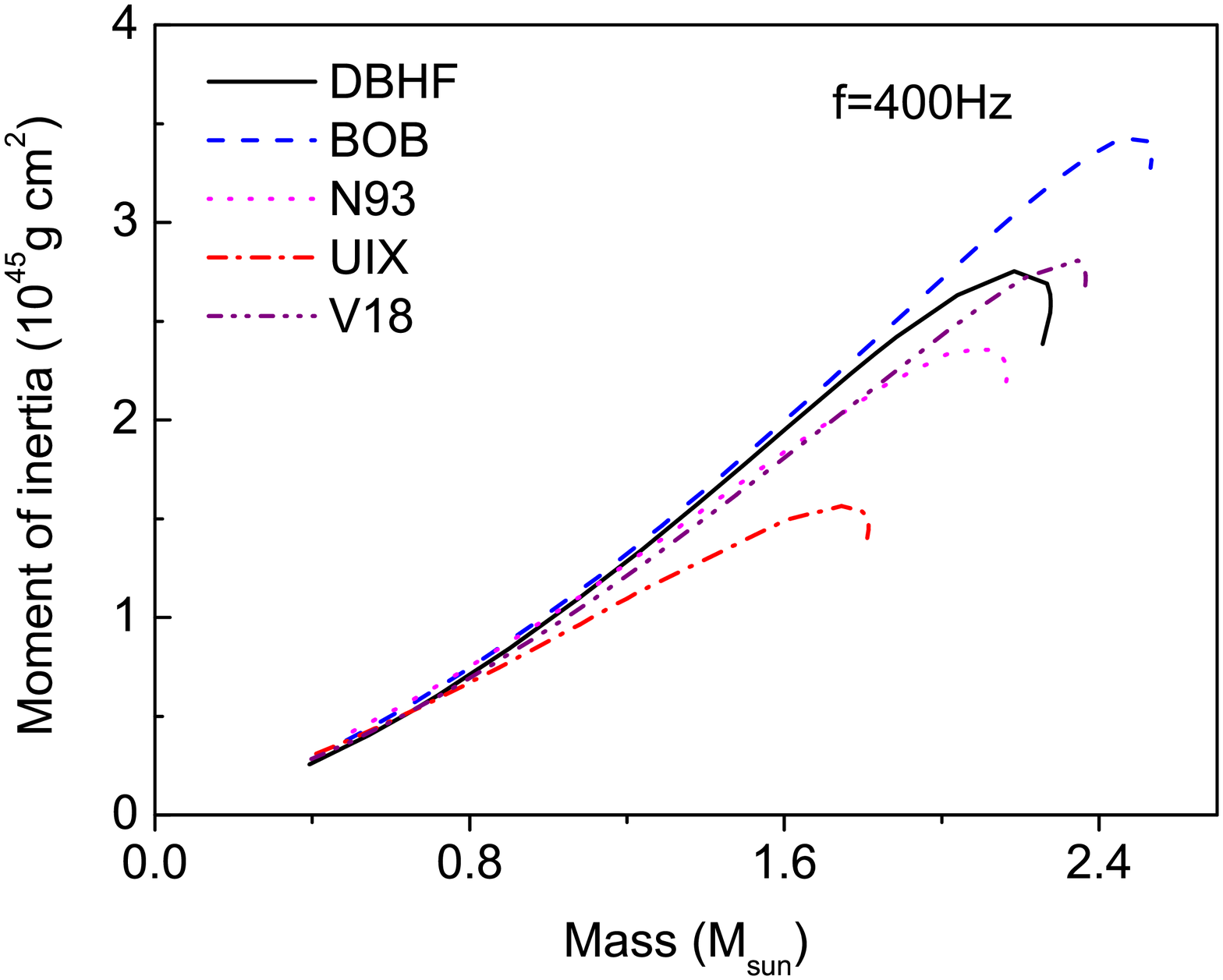}  
\includegraphics[totalheight=1.8in]{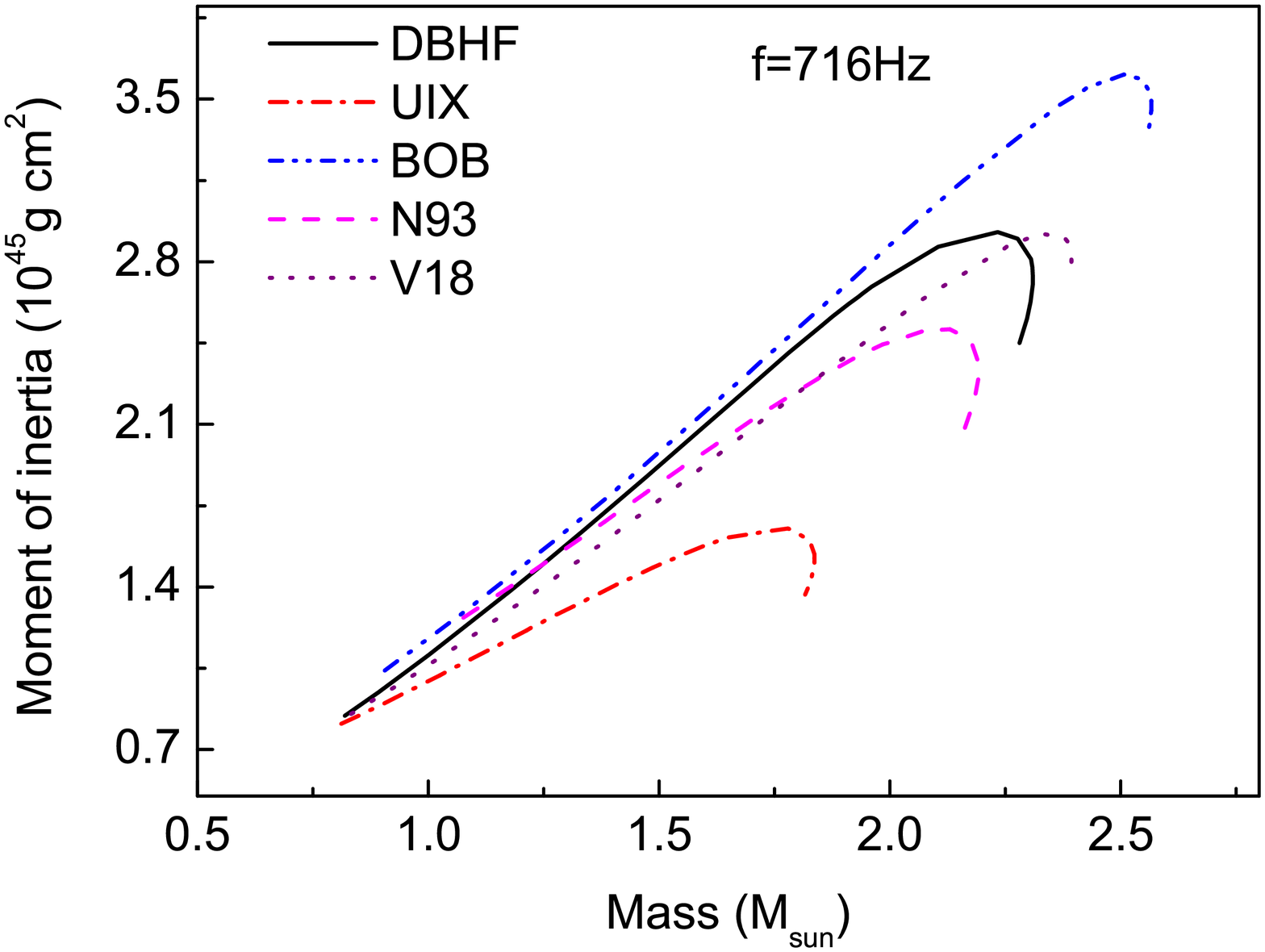}  
\includegraphics[totalheight=1.8in]{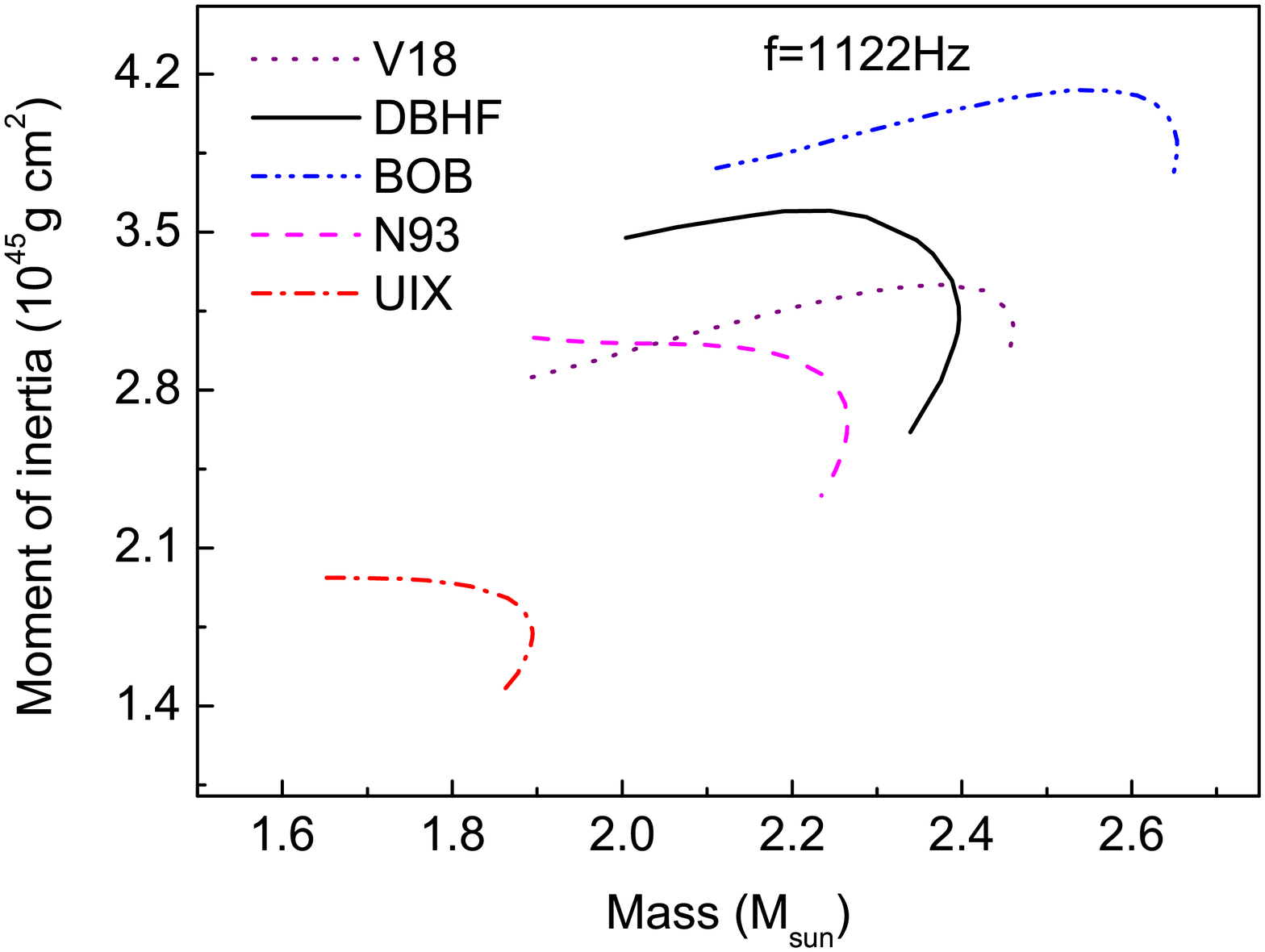}   
\vspace*{-0.5cm}
\caption{(color online)  Moment of inertia for  the models
considered in the text and for different rotational frequencies.                  
} 
\label{six}
\end{figure}

\begin{figure}[!t] 
\centering          
\includegraphics[totalheight=2.5in]{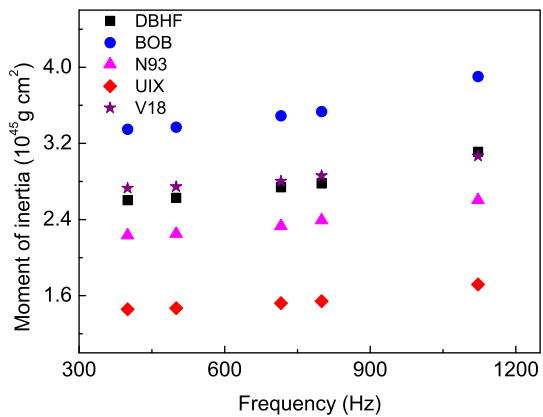}  
\vspace*{-0.5cm}
\caption{(color online)  Moment of inertia corresponding to the maximum mass for  the models
considered in the text as a function of the rotational frequency.                    
} 
\label{seven}
\end{figure}
Lastly, we calculate the gravitational redshift predicted by each model. 
The redshift is defined as 
\begin{equation}
z = \Big (1 - \frac{2M}{R} \Big ) ^{-1/2} -1 \; . 
\end{equation}
This simple formula can be derived considering a photon emitted at the surface of a neutron star and moving
towards a detector located at large distance. The photon frequency at the emitter (receiver) is 
the inverse of the proper time between two wave crests in the frame of the emitter (receiver).   
Assuming a static gravitational field, and writing  $g_{00}$ as the metric tensor component 
at the surface of a nonrotating star yield the equation above. 
Naturally the rotation of the star modifies the metric, and in that case different considerations need to
be applied which result in a frequency dependence of the redshift. We will not be calculating the general case
here. 

From Fig.~8, it appears that the gravitational redshift is not very EoS-dependent (compare, for instance, 
the values at the maximum mass for each model), an indication that the gravitational profile at 
the surface of the star is similar in all models.               
Thus measurements of $z$ may not be the best way to discriminate among EoS's. 

\begin{figure}[!t] 
\centering          
\includegraphics[totalheight=2.5in]{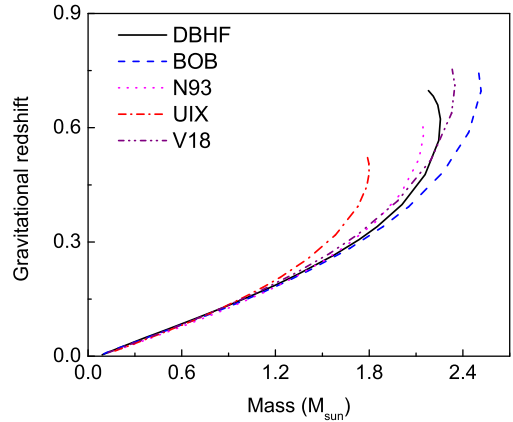}                
\vspace*{-0.5cm}
\caption{(color online)  Gravitational redshift for all models. 
For each model, the corresponding sequence of static stars is considered. 
} 
\label{eight}
\end{figure}

\section{Summary and conclusions}                                                           
We stress the importance of {\it ab initio} calculations to complement the wealth 
of present/future experiments and observations aimed at constraining the less known aspects
of the nuclear equation of state. In that spirit, 
we have undertaken a comparison of EoS-sensitive observables in microscopic models, 
which do or do not include explicit TBF. 
In a previous work \cite{SL09} we related different predictions of the neutron skin in 
$^{208}$Pb to differences in the slope of the symmetry energy. An accurate measurement of the 
neutron skin, as expected to be taken at the Jefferson Laboratory \cite{Jlab}, is the most 
promising way to discriminate among EoS's in the low-density regime. 

At the high-densities           
probed by neutron stars, the model dependence is even larger, however, presently available constraints are 
still insufficient to discriminate among these EoS's.

At high density ($\rho$ between five and ten times normal density), the most repulsive {\it 
symmetric matter} energies are produced with BOB, V18, DBHF, N93 and UIX, in that order.
That is consistent with the maximum mass predictions, which depend mostly on the absolute repulsion
present in the symmetric matter EoS. 
In pure neutron matter, N93 follows right after BOB (again, from largest to smallest repulsion). This indicates
a somewhat different balance when only T=1 contributions are 
included. 
Finally, the symmetry energy, which depends entirely on the repulsion of neutron matter {\it relative} 
to symmetric matter and 
whose density dependence controls observables such as the neutron skin, 
is largest in N93, followed by BOB, V18, UIX, and DBHF.

The model dependence we observe comes from two sources, the two-body potential and the 
many-body approach, specifically the presence of explicit TBF or Dirac effects. 
The dependence on the two-body potential is very large. Typically, the main source of model dependence 
among NN potentials is found in the strength of the tensor force.                                            
Of course, differences at the two-body level impact the TBF as well, whether they are microscopic
or phenomenological (as in the case of UIX). 

On the other hand, 
when comparing DBHF and BOB, we are looking at differences stemming from the many-body scheme, as          
the two models share the same NN potential. 
In the BOB model, repulsion grows at a much faster rate than in DBHF, and more strongly so in neutron 
matter. (Hence, the much larger symmetry energy with BOB \cite{SL09}). As an example, at about 6 times
normal density the DBHF energy of symmetric matter is 67\% of the BOB energy and only 51\% at
ten times normal densities. In pure neutron matter, those ratios become 49\% and 30\%, respectively. 
Thus, the inclusion of the microscopic TBF in the BHF model introduces considerable more 
repulsion than the Dirac effects (whose density dependence is approximately as given in Eq.~(4)), 
through highly non-linear terms            
\cite{cat}.                                      
Both attractive and repulsive TBF are required for a realistic description of the 
saturation point. The density dependence of the repulsive terms is obviously stronger and thus 
dominate at high density. 
Furthermore, it appears that this is especially true in neutron matter. 
The much larger (compared with DBHF) repulsion in neutron matter (relative to symmetric matter) seen in all BHF+TBF models  
may be due to the attractive nature of some T=0 NN partial waves, most noticeably
$^3S_1$, which, on the other hand, would become repulsive at high density in the presence 
of Dirac effects. 
This attractive component (not present in neutron matter) can moderate the repulsion arising from TBF. 

At this time, available constraints cannot pin down the high-density behavior of the EoS. 
Nevertheless, our point is that microscopic models allow for a deeper insight into the            
origin of the observed physical effects, 
and should be pursued along with improved constraints.

\begin{table}
\tiny          
\vspace*{-2cm}
\caption{Equation of state for $\beta$-equilibrated matter. At subnuclear densities the 
crustal EoS from Refs.~\cite{HW,NV} is given. At normal to super-high densities the DBHF
equation of state is shown.} 
\begin{center}
\begin{tabular}{|c|c|c|}
\hline
Baryon density(1/$cm^3$)  & Energy density($g/cm^3$)  & Pressure($dyne/cm^2$)  \\                                         
                  \cline{1-3}
  0.59701000E+25 &  0.99998600E+01 &  0.40721410E+12\\
  0.75334200E+25 &  0.12618390E+02 &  0.15080560E+13\\
  0.11995300E+26 &  0.20092030E+02 &  0.10132810E+14\\
  0.19099900E+26 &  0.31992190E+02 &  0.47174340E+14\\
  0.24101400E+26 &  0.40369630E+02 &  0.94073580E+14\\
  0.30412500E+26 &  0.50940640E+02 &  0.18050450E+15\\
  0.38376300E+26 &  0.64279760E+02 &  0.33550870E+15\\
  0.48425400E+26 &  0.81112010E+02 &  0.60721550E+15\\
  0.61105800E+26 &  0.10235170E+03 &  0.10743490E+16\\
  0.15492600E+27 &  0.25949850E+03 &  0.88776630E+16\\
  0.19549400E+27 &  0.32745010E+03 &  0.14569410E+17\\
  0.24668500E+27 &  0.41319610E+03 &  0.23670880E+17\\
  0.31128200E+27 &  0.52139300E+03 &  0.38114990E+17\\
  0.39279300E+27 &  0.65792350E+03 &  0.60884010E+17\\
  0.49564800E+27 &  0.83020530E+03 &  0.96560320E+17\\
  0.62543700E+27 &  0.10476010E+04 &  0.15215920E+18\\
  0.12566500E+28 &  0.21048780E+04 &  0.57628220E+18\\
  0.20009400E+28 &  0.33515470E+04 &  0.13693280E+19\\
  0.25249000E+28 &  0.42291870E+04 &  0.20992040E+19\\
  0.31860600E+28 &  0.53366300E+04 &  0.32077510E+19\\
  0.40203500E+28 &  0.67340590E+04 &  0.48872650E+19\\
  0.50731100E+28 &  0.84974150E+04 &  0.74261220E+19\\
  0.64015400E+28 &  0.10722520E+05 &  0.11256160E+20\\
  0.12862200E+29 &  0.21544010E+05 &  0.38701870E+20\\
  0.20480200E+29 &  0.34304300E+05 &  0.87378890E+20\\
  0.25843000E+29 &  0.43286960E+05 &  0.13099750E+21\\
  0.32610100E+29 &  0.54622010E+05 &  0.19612730E+21\\
  0.41149300E+29 &  0.68925200E+05 &  0.29327530E+21\\
  0.51924500E+29 &  0.86973760E+05 &  0.43804640E+21\\
  0.65521200E+29 &  0.10974830E+06 &  0.65359850E+21\\
  0.13164600E+30 &  0.22050990E+06 &  0.21592540E+22\\
  0.16611800E+30 &  0.27825210E+06 &  0.32107790E+22\\
  0.20961700E+30 &  0.35111490E+06 &  0.47710750E+22\\
  0.26450500E+30 &  0.44305570E+06 &  0.70850350E+22\\
  0.33376600E+30 &  0.55907310E+06 &  0.10515070E+23\\
  0.42116300E+30 &  0.70547070E+06 &  0.15597180E+23\\
  0.53144400E+30 &  0.89020260E+06 &  0.23123900E+23\\
  0.67060100E+30 &  0.11233090E+07 &  0.33677760E+23\\
  0.17001500E+31 &  0.28479990E+07 &  0.13810890E+24\\
  0.21453100E+31 &  0.35937570E+07 &  0.19523170E+24\\
  0.27070400E+31 &  0.45348070E+07 &  0.27521560E+24\\
  0.34158300E+31 &  0.57222920E+07 &  0.38687930E+24\\
  0.43102100E+31 &  0.72207080E+07 &  0.54232090E+24\\
  0.54387400E+31 &  0.91114890E+07 &  0.75808450E+24\\
  0.68627500E+31 &  0.11497400E+08 &  0.10567510E+25\\
  0.13787700E+32 &  0.23100980E+08 &  0.28166750E+25\\
  0.21952400E+32 &  0.36783270E+08 &  0.53468330E+25\\
  0.27699700E+32 &  0.46415170E+08 &  0.73411590E+25\\
  0.34951600E+32 &  0.58569360E+08 &  0.10057550E+26\\
  0.44101900E+32 &  0.73906140E+08 &  0.13750910E+26\\
  0.55647700E+32 &  0.93259070E+08 &  0.18764540E+26\\
  0.70215700E+32 &  0.11767960E+09 &  0.25560490E+26\\
  0.17797700E+33 &  0.29836060E+09 &  0.86649420E+26\\
  0.22456600E+33 &  0.37648930E+09 &  0.11718980E+27\\
  0.28334500E+33 &  0.47507410E+09 &  0.15831900E+27\\
  0.35751000E+33 &  0.59947540E+09 &  0.21366950E+27\\
  0.45108900E+33 &  0.75645300E+09 &  0.28810670E+27\\
  0.56915000E+33 &  0.95453350E+09 &  0.38815460E+27\\
  0.71811000E+33 &  0.12044860E+10 &  0.52255320E+27\\
  0.14423300E+34 &  0.24200880E+10 &  0.12701340E+28\\
  0.22959600E+34 &  0.38534730E+10 &  0.22901680E+28\\
  0.28966900E+34 &  0.48625310E+10 &  0.30733120E+28\\
  0.36545800E+34 &  0.61358160E+10 &  0.41227540E+28\\
  0.46107400E+34 &  0.77425110E+10 &  0.55287600E+28\\
  0.58169000E+34 &  0.97699500E+10 &  0.74121190E+28\\
\hline
\end{tabular}
\end{center}
\end{table}

\begin{table}
\begin{center}
TABLE I, cont. \\
\vspace*{0.5cm}
\tiny          
\begin{tabular}{|c|c|c|}
\hline
Baryon density(1/$cm^3$)  & Energy density($g/cm^3$)  & Pressure($dyne/cm^2$)  \\                                         
                  \cline{1-3}
  0.73385500E+34 &  0.12328280E+11 &  0.99344590E+28\\
  0.14734400E+35 &  0.24770450E+11 &  0.23888140E+29\\
  0.18587600E+35 &  0.31256660E+11 &  0.31991950E+29\\
  0.23448300E+35 &  0.39441580E+11 &  0.42838370E+29\\
  0.29580200E+35 &  0.49769610E+11 &  0.57354730E+29\\
  0.37313900E+35 &  0.62801930E+11 &  0.76780650E+29\\
  0.47069400E+35 &  0.79247170E+11 &  0.10277560E+30\\
  0.59374600E+35 &  0.99998600E+11 &  0.13755860E+30\\
  0.10899500E+36 &  0.18374080E+12 &  0.15080040E+30\\
  0.20791800E+36 &  0.35086530E+12 &  0.41327350E+30\\
  0.30685800E+36 &  0.51813430E+12 &  0.66738040E+30\\
  0.40581200E+36 &  0.68549240E+12 &  0.89490400E+30\\
  0.50477400E+36 &  0.85290930E+12 &  0.10966500E+31\\
  0.60373600E+36 &  0.10203670E+13 &  0.12773240E+31\\
  0.80166500E+36 &  0.13553720E+13 &  0.15941520E+31\\
  0.90062000E+36 &  0.15229090E+13 &  0.17377690E+31\\
  0.10985200E+37 &  0.18580330E+13 &  0.20087290E+31\\
  0.11974500E+37 &  0.20256030E+13 &  0.21397550E+31\\
  0.13953200E+37 &  0.23608150E+13 &  0.23994040E+31\\
  0.15931600E+37 &  0.26960450E+13 &  0.26615840E+31\\
  0.16920800E+37 &  0.28636860E+13 &  0.27951740E+31\\
  0.17910000E+37 &  0.30313280E+13 &  0.29309910E+31\\
  0.19888300E+37 &  0.33666280E+13 &  0.32106350E+31\\
  0.20877300E+37 &  0.35342880E+13 &  0.33549110E+31\\
  0.21866300E+37 &  0.37019650E+13 &  0.35024070E+31\\
  0.23844100E+37 &  0.40373370E+13 &  0.38075420E+31\\
  0.24833000E+37 &  0.42050320E+13 &  0.39653560E+31\\
  0.25821800E+37 &  0.43727450E+13 &  0.41267750E+31\\
  0.26810600E+37 &  0.45404400E+13 &  0.42918160E+31\\
  0.27799300E+37 &  0.47081710E+13 &  0.44605250E+31\\
  0.28788000E+37 &  0.48758840E+13 &  0.46329190E+31\\
  0.29776600E+37 &  0.50436140E+13 &  0.48089980E+31\\
  0.30765200E+37 &  0.52113630E+13 &  0.49887950E+31\\
  0.31753700E+37 &  0.53790930E+13 &  0.51722760E+31\\
  0.32742200E+37 &  0.55468420E+13 &  0.53594420E+31\\
  0.33730600E+37 &  0.57146080E+13 &  0.55502940E+31\\
  0.34718900E+37 &  0.58823750E+13 &  0.57447980E+31\\
  0.35707300E+37 &  0.60501410E+13 &  0.59429390E+31\\
  0.36695600E+37 &  0.62179070E+13 &  0.61447010E+31\\
  0.37683800E+37 &  0.63856910E+13 &  0.63500680E+31\\
  0.38672000E+37 &  0.65534760E+13 &  0.65589600E+31\\
  0.39660200E+37 &  0.67212770E+13 &  0.67714090E+31\\
  0.40648400E+37 &  0.68890790E+13 &  0.69873660E+31\\
  0.41636500E+37 &  0.70568810E+13 &  0.72068480E+31\\
  0.42624500E+37 &  0.72246830E+13 &  0.74297430E+31\\
  0.43612600E+37 &  0.73925030E+13 &  0.76560670E+31\\
  0.44600500E+37 &  0.75603230E+13 &  0.78857870E+31\\
  0.46576400E+37 &  0.78959800E+13 &  0.83552890E+31\\
  0.47564200E+37 &  0.80638360E+13 &  0.85949750E+31\\
  0.48552100E+37 &  0.82316740E+13 &  0.88379610E+31\\
  0.49539800E+37 &  0.83995290E+13 &  0.90841830E+31\\
  0.50527400E+37 &  0.85673840E+13 &  0.93336260E+31\\
  0.51515000E+37 &  0.87352400E+13 &  0.95862580E+31\\
  0.52502600E+37 &  0.89031130E+13 &  0.98420450E+31\\
  0.53490100E+37 &  0.90709860E+13 &  0.10100920E+32\\
  0.54477600E+37 &  0.92388600E+13 &  0.10362880E+32\\
  0.55465100E+37 &  0.94067330E+13 &  0.10627910E+32\\
  0.56452500E+37 &  0.95746240E+13 &  0.10896010E+32\\
  0.57439900E+37 &  0.97425150E+13 &  0.11167080E+32\\
  0.58427200E+37 &  0.99104240E+13 &  0.11441110E+32\\
  0.59414500E+37 &  0.10078320E+14 &  0.11718180E+32\\
  0.60401800E+37 &  0.10246220E+14 &  0.11998160E+32\\
  0.61389000E+37 &  0.10414150E+14 &  0.12280980E+32\\
  0.63363400E+37 &  0.10749990E+14 &  0.12855310E+32\\
  0.65337700E+37 &  0.11085860E+14 &  0.13440860E+32\\
\hline
\end{tabular}
\end{center}
\end{table}
\begin{table}
\begin{center}
TABLE I, cont. \\ 
\vspace*{0.5cm}
\tiny              
\begin{tabular}{|c|c|c|}
\hline
Baryon density(1/$cm^3$)  & Energy density($g/cm^3$)  & Pressure($dyne/cm^2$)  \\                                         
                  \cline{1-3}
  0.66324800E+37 &  0.11253780E+14 &  0.13737910E+32\\
  0.67311800E+37 &  0.11421730E+14 &  0.14037590E+32\\
  0.69285800E+37 &  0.11757640E+14 &  0.14645200E+32\\
  0.70272800E+37 &  0.11925600E+14 &  0.14953050E+32\\
  0.72246600E+37 &  0.12261520E+14 &  0.15576620E+32\\
  0.73233400E+37 &  0.12429500E+14 &  0.15892440E+32\\
  0.75207000E+37 &  0.12765460E+14 &  0.16531580E+32\\
  0.76193800E+37 &  0.12933460E+14 &  0.16855220E+32\\
  0.77180500E+37 &  0.13101440E+14 &  0.17181270E+32\\
  0.79153800E+37 &  0.13437440E+14 &  0.17840720E+32\\
  0.80140400E+37 &  0.13605450E+14 &  0.18174290E+32\\
  0.82113400E+37 &  0.13941490E+14 &  0.18848650E+32\\
  0.83099900E+37 &  0.14109500E+14 &  0.19189430E+32\\
  0.85072800E+37 &  0.14445550E+14 &  0.19878210E+32\\
  0.86059200E+37 &  0.14613590E+14 &  0.20226200E+32\\
  0.87045500E+37 &  0.14781620E+14 &  0.20576440E+32\\
  0.88031800E+37 &  0.14949670E+14 &  0.20929080E+32\\
  0.89018100E+37 &  0.15117700E+14 &  0.21284120E+32\\
  0.90990300E+37 &  0.15453810E+14 &  0.22001090E+32\\
  0.91976300E+37 &  0.15621880E+14 &  0.22362710E+32\\
  0.93948300E+37 &  0.15958020E+14 &  0.23093300E+32\\
  0.94934200E+37 &  0.16126090E+14 &  0.23461640E+32\\
  0.98877600E+37 &  0.16798420E+14 &  0.24958070E+32\\
  0.14590200E+38 &  0.24523600E+14 &  0.41632200E+32\\
  0.23168800E+38 &  0.38982700E+14 &  0.82280400E+32\\
  0.34584300E+38 &  0.58250100E+14 &  0.14210400E+33\\
  0.49242100E+38 &  0.83026900E+14 &  0.23006100E+33\\
  0.67547500E+38 &  0.11401900E+15 &  0.40981400E+33\\
  0.78194600E+38 &  0.13207300E+15 &  0.54012600E+33\\
  0.89905700E+38 &  0.15195500E+15 &  0.71172400E+33\\
  0.10273100E+39 &  0.17376500E+15 &  0.97049800E+33\\
  0.11672200E+39 &  0.19760500E+15 &  0.13704600E+34\\
  0.13192900E+39 &  0.22358900E+15 &  0.20162700E+34\\
  0.14840200E+39 &  0.25183700E+15 &  0.26192800E+34\\
  0.16619200E+39 &  0.28248100E+15 &  0.33029700E+34\\
  0.18535000E+39 &  0.31568300E+15 &  0.51720500E+34\\
  0.20592700E+39 &  0.35161900E+15 &  0.75702800E+34\\
  0.22797300E+39 &  0.39049500E+15 &  0.11003700E+35\\
  0.25153800E+39 &  0.43255500E+15 &  0.16084300E+35\\
  0.27667400E+39 &  0.47813800E+15 &  0.23509200E+35\\
  0.30343200E+39 &  0.52760100E+15 &  0.33455700E+35\\
  0.33186100E+39 &  0.58135500E+15 &  0.46583400E+35\\
  0.36201200E+39 &  0.63986100E+15 &  0.63214300E+35\\
  0.39393700E+39 &  0.70359400E+15 &  0.83222800E+35\\
  0.42768500E+39 &  0.77303000E+15 &  0.10713900E+36\\
  0.46330800E+39 &  0.84870300E+15 &  0.13477500E+36\\
  0.50085600E+39 &  0.93108800E+15 &  0.16560600E+36\\
  0.54038000E+39 &  0.10206900E+16 &  0.20027100E+36\\
  0.58193000E+39 &  0.11181400E+16 &  0.24141700E+36\\
  0.62555700E+39 &  0.12243400E+16 &  0.29312500E+36\\
  0.67131200E+39 &  0.13404700E+16 &  0.35869100E+36\\
  0.71924500E+39 &  0.14679700E+16 &  0.44217300E+36\\
  0.76940800E+39 &  0.16084100E+16 &  0.53909800E+36\\
  0.82185000E+39 &  0.17629800E+16 &  0.64928300E+36\\
  0.87662200E+39 &  0.19333700E+16 &  0.77966700E+36\\
  0.93377600E+39 &  0.21214100E+16 &  0.93298700E+36\\
  0.99336100E+39 &  0.23293000E+16 &  0.11139500E+37\\
  0.10554300E+40 &  0.25594100E+16 &  0.13245200E+37\\
  0.11200300E+40 &  0.28143700E+16 &  0.15691600E+37\\
  0.11872100E+40 &  0.30972100E+16 &  0.18569700E+37\\
  0.12570300E+40 &  0.34114400E+16 &  0.21904600E+37\\
  0.13295400E+40 &  0.37606000E+16 &  0.25722700E+37\\
  0.14047800E+40 &  0.41488100E+16 &  0.30137700E+37\\
  0.14828000E+40 &  0.45806700E+16 &  0.35220200E+37\\
  0.15636600E+40 &  0.50613000E+16 &  0.41065500E+37\\
  0.16474100E+40 &  0.55963600E+16 &  0.47754300E+37\\
  0.17341000E+40 &  0.61920200E+16 &  0.55414900E+37\\
  0.18237800E+40 &  0.68552200E+16 &  0.64123300E+37\\
\hline
\end{tabular}
\end{center}
\end{table}
\section{Acknowledgments}                                                           
Support from the U.S. Department of Energy under Grant No. DE-FG02-03ER41270 is 
acknowledged.                                                                          
We are grateful to F. Weber for providing          
the crustal EoS.

\end{document}